\documentclass[12pt]{article}
\usepackage[english,russian]{babel}
\usepackage{epsfig}
\begin{document}
\large

\par
\noindent {\bf Neutrino oscillations in the scheme of mass mixings
and problem of smallness of angle mixing $\theta_{1 3}$}

\par
\begin{center}
\vspace{0.3cm} Kh. M. Beshtoev
\par
\vspace{0.3cm} Joint Institute for Nuclear Research, Joliot Curie
6, 141980 Dubna, Moscow region, Russia.
\end{center}
\vspace{0.3cm}

\par
Abstract\\

\par
In the framework of the mass mixing scheme we have considered
mixings and oscillations of $\nu_e, \nu_\mu, \nu_\tau$ neutrinos
and obtained expressions for angle mixings and lengths of
oscillations in dependence on components of the nondiagonal mass
matrix. Then analysis of these obtained results was done by using
modern experimental data on neutrino oscillations. It has been
shown that in this approach the lengths of neutrino oscillations
$L_{2 3}$ and $L_{1 3}$ are not compulsory to be equal. It means
that the angle mixing $\theta_{1 3}$ can be not very small, i.e.,
$L_{1 3}$ can be larger than $L_{2 3}$.
\par
In the conventional approach $L_{1 3} \approx L_{2 3}$ ($L_{1 2}
\gg L_{2 3}$) and angle mixing of $\theta_{1 3}$ is very small.
Angle mixings $\theta_{2 3}, \theta_{1 2}$ are big. Then there ia
a problem: why is mixing angle $\theta_{1 3}$ so small? A natural
solution of the problem is to suppose that $(m_2^2 - m_1^2) \neq
(m_3^2 - m_1^2) - (m_3^2 - m_2^2)$, then $L_{1 3} > L_{2 3}$. It
will be realized if there are 4 neutrino oscillations instead of 3
neutrino oscillations. Then the value of $\theta_{1 3}$ is
necessary to search at distances more than $L_{2 3}$.


\section{\bf Introduction}

\par
Oscillations of $K^o$ mesons (i.e., $K^{o} \leftrightarrow \bar
K^{o}$) were theoretically \cite{1} and experimentally \cite{2}
investigated in the 50-s and 60-s. Recently an understanding has
been achieved that these processes go as a double-stage process
\cite{3,4,5,6}. A detailed study of $К^{о}$ meson mixing and
oscillations is very important since the theory of neutrino
oscillations is built in analogy with the theory of $К^{о}$ meson
oscillations.
\par
The suggestion that, in analogy with $K^{o},\bar K^{o}$
oscillations, there could be neutrino-antineutrino oscillations (
$\nu \rightarrow \bar \nu$), was considered by Pontecorvo \cite{7}
in 1957. It was subsequently considered by Maki et al. \cite{8}
and Pontecorvo \cite{9} that there could be mixings (and
oscillations) of neutrinos of different flavors (i.e., $\nu _{e}
\rightarrow \nu _{\mu }$ transitions).
\par
Lengths of three neutrino oscillations are \cite{10} as follows
$$
L_{1 2} = 2\pi \frac{2 p}{\mid m_2^2 - m_1^2 \mid}, \quad L_{1 3}
= 2\pi \frac{2 p}{\mid m_3^2 - m_1^2 \mid}, \quad L_{2 3} = 2\pi
\frac{2 p}{\mid m_3^2 - m_2^2 \mid}. \eqno(1)
$$
In experiments \cite{11, 12} $L_{1 2}$ and $L_{2 3}$ were measured
and it was obtained that $L_{1 2} \gg L_{2 3}$. If to use the
expression
$$
(m_2^2 - m_1^2) = (m_3^2 - m_1^2) - (m_3^2 - m_2^2), \eqno(2)
$$
taking into account that $m_2^2 - m_1^2 = \frac{4 \pi p}{L_{1
2}}$, $m_3^2 - m_1^2 = \frac{4 \pi p}{L_{1 3}}$, $m_3^2 - m_2^2 =
\frac{4 \pi p}{L_{2 3}}$, then we obtain
$$
L_{1 3} = \frac{L_{1 2} L_{2 3}}{L_{1 2} + L_{2 3}}. \eqno(3)
$$
Since $L_{1 2} \gg L_{2 3}$, then from expression (3) we get
$$
L_{1 3} \approx L_{2 3}. \eqno(4)
$$
The mixing angle $\theta_{1 3}$ measured in experiment \cite{11}
is very small. Then there is a question:  why are mixing angles
$\theta_{2 3}$ and $\theta_{1 2}$ measured in the experiment of
the order of maximal angle ($\pi/4$) and why is the measured angle
$\theta_{1 3}$ is so small? There is definitely a problem. To
investigate this problem, we will work in the framework of the
masses mixing scheme.
\par
Since the scheme (theory) of neutrino oscillations is constructed
in analogy with $K^o$ mesons, at first we consider the scheme of
$K^o$ meson oscillations and then the scheme of neutrino
oscillations.

\section{Vacuum mixings and oscillations of $K^o, \bar K^o$ mesons at
strangeness violation by weak interactions}

\par
$K^{o}$-, $\bar K^{o}$- meson states are produced in the strong
interaction (i.e, they are eigenstates of these interactions),
then the mass matrix of $K^{o}$ - mesons will have a diagonal form
\cite{3, 4, 5, 6}. Following the traditions we will consider the
$K^{o}$ - meson mixings and oscillations by using the mass matrix
and for convenience the masses are used in the linear but not in
the quadratic form, then the mass matrix has the following form:
$$
\left(\begin{array}{cc} m_{K^o} & 0 \\ 0 & m_{\bar K^0}
\end{array} \right) .
\eqno(5)
$$
\par
Because of the weak interactions violating strangeness $(s
\leftrightarrow d)$ this mass matrix (5) becomes a nondiagonal
matrix:
$$
\left(\begin{array}{cc} m_{K^o} & m_{K^o \bar K^o} \\
m_{\bar K^o K^o}  & m_{\bar K^o}
\end{array} \right). \eqno(6)
$$
\par
For obtaining the eigenstates of weak interactions which violate
strangeness, we have to diagonalize this matrix by turning on
angle $\theta_o$:
$$
\left(\begin{array}{cc} m_{K^o} & m_{K^o \bar K^o} \\
m_{\bar K^o K^o}  & m_{\bar K^o}
\end{array} \right) \to U^-1 \left(\begin{array}{cc} m_{K^o_1} & 0 \\
0  & m_{K^o_2} \end{array} \right) U , U = \left(\begin{array}{cc} cos\theta_o & -sin\theta_o \\
sin\theta_o & cos\theta_o \end{array} \right) .  \eqno(7)
$$
By using this procedure, we get
$$
tg 2\theta_o = \frac{2m_{K^o \bar K^o}}{\mid m_{K^o} - m_{\bar
K^o} \mid}   ,
$$
$$
sin 2\theta_o = \frac{2m_{K^o \bar K^o}}{\sqrt{(m_{K^o} - m_{\bar
K^o})^2 +(2m_{K^o \bar K^o})^2}}  ,  \eqno(8)
$$
$$
m_{1, 2} =  m_{K_1, K_2} = {1\over 2} \left[ (m_{K^o} + m_{\bar
K^o}) \pm \left((m_{K^o} - m_{\bar K^o})^2 + 4 m^{2}_{K^o \bar
K^o} \right)^{1/2} \right]  , \eqno(9)
$$
where $K^o_1$ and $K^o_2$ states are eigenstates of the weak
interactions violating strangeness. Now these states are
superposition states of $K^o, \bar K^o$ mesons
$$
\begin{array} {cc} K^o_1 = {cos \theta_o K^o - sin \theta_o \bar K^o} , \\ K^o_2
= { sin \theta_o K^o + cos \theta_o \bar K^o}, \end{array}
\eqno(10)
$$
and the inverse transformation gives the following:
$$
\begin{array} {cc} K^o = {cos \theta_o K^o_1 + sin \theta_o K^o_2} , \\ \bar K^o
= {- sin \theta_o K^o_1 + cos \theta_o \bar K^o_2} .\end{array}
\eqno(11)
$$
\par
The computation of nondiagonal components of the mass matrix (6)
can be fulfilled by using the Feynman block diagrams in the
framework of the standard model of electroweak interactions
\cite{13}, \cite{14} with Kabibbo-Kobayashi-Maskawa \cite{15}
matrices.
\par
Expression for $sin^2 2\theta_o$ can be obtained from (8) and it
is given by the following expression ($\theta_o$ is the angle of
mixing):
$$
sin^2 2\theta_o = \frac{(2m_{K^o \bar K^o})^2} {(m_{K^o} - m_{\bar
K^o})^2 +(2m_{K^o \bar K^o})^2} . \eqno(12)
$$
This expression has the analogy with a well knowing formula of
Breit-Wigner \cite{16} for transition probability,
$$
W(m , m_{o}, \Gamma, \ldots) = {{c ({\Gamma \over 2})^2} \over {(m
- m_o)^2 + ({\Gamma \over 2})^2}} , \eqno(13)
$$
where $c$ - is a normalized factor, $m, m_o$ are masses and
$\Gamma$ is the width transition. We can obtain expression (12)
from Breit-Wigner formula if we fulfill the following
substitutions:
$$
m \to m_{K^o}, \quad m_o \to m_{\bar K^o}, \quad {\Gamma \over 2}
\to 2m_{K^o \bar K^o} \quad c \to 1 . \eqno(14)
$$
We see that we can interpret nondiagonal mass term $2m_{K^o \bar
K^o}$ as half-width of $K^o \leftrightarrow \bar K^o$ transitions.
\par
Since the weak interactions are $CPT$ invariant, then $m_{K^{o}
K^{o}} = m_{\bar K^o \bar K^o}$ and the mixing angle $\theta_o$
will be equal to ${\pi \over 4}$, i.e., $sin^2 2 \theta_o = 1$.
Then from expression (14) and (11) we get
$$
K^o_1 = {{K^o - \bar K^o}\over {\sqrt{2}}} , \qquad K^o_2 = {{K^o
+ \bar K^o}\over {\sqrt{2}}} , \eqno(15)
$$
$$
K^o = \frac{K^o_1 +  K^o_2}{\sqrt{2}}, \qquad \bar K^o =
\frac{K^o_1 - K^o_2}{\sqrt{2}} . \eqno(15')
$$
It is necessary to remark that $CP$$K^o_1 = K^o_1$ and $CP K^o_2 =
-K^o_2$, i.e., $CP$ parity $K^o_1$ meson is a positive value and
it can decay into two $\pi$ mesons, and $CP$ parity of $K^o_2$
meson is a negative value and it can decay into three $\pi$
mesons.
\par
The evolution of $K^o _{1}, K^o _{2}$ meson states with masses
$m_{1}, m_{2}$ will be given with the following expression:
$$
K^o _{1}(t) = e^{-i E_1 t} K^o _{1}(0), \qquad K^o _{2}(t) = e^{-i
E_2 t} K^o _{2}(0) , \eqno(16)
$$
where
$$
E^2_{k} = (p^{2} + m^2_{k}), k = 1, 2  .
$$
\par
If these mesons are moving without interactions, then
$$
\begin{array}{c}
K^o(t) = cos \theta_o e^{-i E_1 t} K^o_{1}(0) + sin \theta_o
e^{-i E_2 t} K^o_{2}(0) , \\
\bar K^o(t) = - sin \theta_o e^{-i E_1 t} K^o_{1}(0) + cos
\theta_o e^{-i E_2 t} K^o_{2}(0) .
\end{array}
\eqno(17)
$$
Using expression (10) for $K^o _{1}$ ? $K^o _{2}$ and putting them
in (17), we obtain
$$
K^o (t) = \left[e^{-i E_1 t} cos^{2} \theta_o + e^{-i E_2 t}
sin^{2} \theta_o \right] K^o(0) +
$$
$$
+ \left[e^{-i E_1 t} - e^{-i E_2 t} \right] sin \theta_o \cos
\theta_o \bar K^o(0) , \eqno(18)
$$
$$
\bar K^o(t) = \left[e^{-i E_1 t} sin^{2} \theta_o + e^{-i E_2 t}
cos^{2} \theta_o \right] \bar K^o(0)  +
$$
$$
+ \left[e^{-i E_1 t} - e^{-i E_2 t} \right] sin\theta_o cos
\theta_o \bar K^o(0) .
$$
\par
The probability that meson $K^o$ produced at moment $t = 0$ will
be at moment $t \neq 0$ in the state of $\bar K^o$ meson is given
by a squared absolute value of the amplitude in (18)
$$
\begin{array}{c}
P(K^o \rightarrow \bar K^o) = \mid(\bar K^o(0) \cdot K^o(t))\mid^2 =\\
 = {1\over 2} \sin^{2} 2\theta_o \left[1 - cos ((E_2 - E_1) t)
\right] \equiv \frac{1}{2} \left[1 - cos ((E_2 - E_1) t) \right] ,
\end{array}, \eqno(19)
$$
where $\theta_o = \pi/4$. Using expressions for masses of $K^o_1,
K^o_2$ mesons, we obtain
$$
m_{K^o_1} = m_{K^o} -  \Delta,  m_{K^o_2} = m_{K^o} +  \Delta,
\eqno(20)
$$
where $\Delta = 2 m_{K^o \bar K^o}$. Since $m_{K^o} \gg \Delta$,
$$
\begin{array}{cc} E_1 = \sqrt{p^2 + m^2_{K^o_1}} \cong E_{K^o} (1 - \frac{m_{K^o}
\Delta}{E_{K^o}^2}), \\ E_2 = \sqrt{p^2 + m^2_{K^o_2}} \cong
E_{K^o} (1 + \frac{m_{K^o} \Delta}{E_{K^o}^2}), \end{array}
\eqno(21)
$$
$$
E_2 - E_1 = \frac{2 m_{K^o} \Delta}{E_{K^o}} = \frac{2
\Delta}{\gamma} , \eqno(22)
$$
Then the length $L_{1 2}$ of $K^o, \bar K^o$ meson oscillations is
a follows:
$$
L_{1 2} = \frac{\gamma}{2 \Delta} \equiv \frac{2 \pi h c \gamma}{2
\Delta} . \eqno(23)
$$
The mixing angle $\theta_o$ of $K^o, \bar K^o$ is equal to $\pi/4$
and the value for $\Delta$ computed in the framework of weak
interactions (see references in \cite{13, 14}) is in a reasonable
agreement with the same value obtained in experiments. So, we see
that the scheme of mass mixings is in a rather good agreement with
the experiment.
\par
Now let as consider neutrino oscillations in the framework of the
mass mixings scheme.

\section{Vacuum mixings and oscillations of $\nu_e, \nu_\mu, \nu_\tau$ neut\-rinos in the scheme of mass mixings}

As it is mentioned above we will work in the mass mixings scheme.
We can use the $3 \times 3$ mass matrix but since the lengths of
neutrino oscillations are noticeably differ, then it is proper to
work by using three $2 \times 2$ mass matrices corresponding to
$\nu_e \to \nu_\mu$, $\nu_\mu \to \nu_\tau$ and $\nu_e \to
\nu_\tau$ mixings and oscillations. Since neutrino oscillations
are considered in analogy with $K^o$ meson oscillations, then we
will use the method analogous to the one considered above.

\subsection{The case of $\nu_e, \nu_\mu$ neutrino oscillations}

\par
If the $\nu_e, \nu_\mu$ neutrino states are produced in the weak
interactions (i.e, they are eigenstates of these interactions,
then the mass matrix of $\nu_e, \nu_\mu$ - neutrinos will have a
diagonal form. Following the traditions we will consider the
$\nu_e, \nu_\mu$ - neutrino mixings and oscillations by using the
mass matrix and for convenience the masses are used in the linear
but not in the quadratic form, then the mass matrix has the
following form:
$$
\left(\begin{array}{cc} m_{\nu_e} & 0 \\ 0 & m_{\nu_\mu}
\end{array} \right) .
\eqno(24)
$$
\par
Since there is a interaction violating lepton numbers this mass
matrix (24) becomes a nondiagonal matrix:
$$
\left(\begin{array}{cc} m_{\nu_e} & m_{\nu_e \nu_\mu} \\
m_{\nu_e \nu_\mu}  & m_{\nu_\mu}
\end{array} \right) .  \eqno(25)
$$
\par
For obtaining the eigenstates of weak interactions which violate
lepton numbers, we have to diagonalize this matrix by turning on
angle $\theta$ ($\theta \equiv \theta_{1 2})$. By using this
procedure, we get
$$
U^{-1} \left(\begin{array}{cc} m_{\nu_e} & m_{\nu_e \nu_\mu} \\
m_{\nu_e \nu_\mu}  & m_{\nu_\mu} \end{array} \right) U
= \left(\begin{array}{cc} m_{\nu_1} & 0 \\
0  & m_{\nu_2} \end{array} \right), \quad U = \left(\begin{array}{cc} cos\theta & -sin\theta \\
sin\theta & cos\theta \end{array} \right) .
$$
$$
tg 2\theta = \frac{2m_{\nu_e \nu_\mu}}{\mid m_{\nu_e} -
m_{\nu_\mu} \mid}   ,
$$
$$
sin 2\theta = \frac{2m_{\nu_e \nu_\mu}}{\sqrt{(m_{\nu_e} -
m_{\nu_\mu})^2 +(2m_{\nu_e \nu_\mu})^2}}  ,  \eqno(26)
$$
$$
m_{1, 2} \equiv  m_{\nu_1 \nu_2} = {1\over 2} \left[ (m_{\nu_e} +
m_{\nu_\mu}) \pm \left((m_{\nu_e} - m_{\nu_\mu})^2 + 4
m^{2}_{\nu_e \nu_\mu} \right)^{1/2} \right]  , \eqno(27)
$$
where $\nu_1$ and $\nu_2$ states are eigenstates of the weak
interactions violating lepton numbers. Now these states are
superposition states of $\nu_e, \nu_\mu$ neutrinos.
$$
\begin{array} {cc} \nu_1 = {cos \theta \nu_e - sin \theta \nu_\mu} , \\ \nu_2
= { sin \theta \nu_e + cos \theta \nu_\mu}, \end{array} \eqno(28)
$$
and the inverse transformation gives:
$$
\begin{array} {cc} \nu_e = {cos \theta \nu_1 + sin \theta \nu_2} , \\ \nu_\mu
= {- sin \theta \nu_1 + cos \theta \nu_2} ,\end{array} \eqno(29)
$$
Then $\nu_e, \nu_\mu$ neutrino masses are connected with masses of
$\nu_1, \nu_2$ neutrinos via the following expressions
$$
\begin{array}{c}
m_{\nu_e} = m_1 {\rm cos}^2 \,\theta + m_2 \sin^2 i\theta,\\
m_{\nu_\mu} = m_1 \sin^2\, \theta + m_2 {cos}^2 \,\theta,
\end{array}
\eqno(30)
$$
Now mass Lagrangian of two neutrinos ($\nu_e, \nu_\mu$) has the
following form:
$$
\begin{array}{c}{\cal L}_{M} = -  \frac{1}{2} \left[m_{\nu_e}
\bar \nu_e \nu_e + m_{\nu_\mu} \bar \nu_{\mu} \nu_{\mu } +
m_{\nu_e \nu_{\mu }}(\bar \nu_e \nu_{\mu } + \bar \nu_{\mu }
\nu _e) \right] \equiv \\
\equiv  -  \frac{1}{2} (\bar \nu_e, \bar \nu_\mu)
\left(\begin{array}{cc} m_{\nu_e} & m_{\nu_e \nu_{\mu }} \\
m_{\nu_{\mu} \nu_e} & m_{\nu_\mu} \end{array} \right)
\left(\begin{array}{c} \nu_e \\ \nu_{\mu } \end{array} \right),
\end{array}
\eqno(31)
$$
while diagonalizing it transforms into the following one:
$$
{\cal L}_{M} = -  \frac{1}{2} \left[ m_{1} \bar \nu_{1} \nu_{1} +
m_{2} \bar \nu_{2} \nu_{2} \right], \eqno(32)
$$
\par
Expression for $sin^2 2\theta$ can be obtained from (30) and it is
given by the following expression ($\theta$ is the vacuum angle of
mixing):
$$
\sin^2\, (2 \theta) = \frac{(2m_{\nu_{e} \nu_{\mu}})^2}
{(m_{\nu_e} - m_{\nu_\mu})^2 +(2m_{\nu_e \nu_{\mu}})^2}. \eqno(33)
$$
This expression has the analogy with a well known formula of
Breit-Wigner \cite{16} for the transition probability:
$$
W(m , m_{o}, \Gamma, \ldots) = {{c ({\Gamma \over 2})^2} \over {(m
- m_o)^2 + ({\Gamma \over 2})^2}} , \eqno(34)
$$
where $c$ - is a normalized factor, $m, m_o$ are masses and
$\Gamma$ is the width transition. We can obtain expression (33)
from Breit-Wigner formula if we fulfill the following
substitutions
$$
m \to m_{\nu_e}, \quad m_o \to m_{\nu_\mu}, \quad {\Gamma \over 2}
\to 2m_{\nu_e \nu_\mu} \quad c \to 1 . \eqno(35)
$$
We see that nondiagonal mass term $2m_{\nu_e \nu_\mu}$ can be
interpreted as half-width of $\nu_e \leftrightarrow \nu_\mu$
transitions.
\par
The evolution of $\nu_1, \nu_2$ neutrino states with masses $m_1,
m_2$ will be given with the following expressions:
$$
\nu_1(t) = e^{-i E_1 t} \nu_1(0), \qquad \nu_2(t) = e^{-i E_2 t}
\nu_2(0) , \eqno(36)
$$
where $E^2_{k} = (p^{2} + m^2_{k})$, $k = 1, 2$.
\par
If these neutrinos are moving without interactions, then
$$
\begin{array}{c}
\nu_e(t) = cos \theta e^{-i E_1 t} \nu_1(0) + sin \theta
e^{-i E_2 t} \nu_2(0) , \\
\nu_\mu(t) = - sin \theta e^{-i E_1 t} \nu_1(0) + cos \theta e^{-i
E_2 t} \nu_2(0) .
\end{array}
\eqno(37)
$$
Using expression (36) for $\nu_1$ and $\nu_2$ and putting them in
(37), we obtain
$$
\nu_e(t) = \left[e^{-i E_1 t} cos^{2} \theta + e^{-i E_2 t}
sin^{2} \theta \right] \nu_e(0) +
$$
$$
+ \left[e^{-i E_1 t} - e^{-i E_2 t} \right] sin \theta \cos \theta
\nu_\mu(0) , \eqno(38)
$$
$$
\nu_\mu (t) = \left[e^{-i E_1 t} sin^{2} \theta + e^{-i E_2 t}
cos^{2} \theta \right] \nu_e (0)  +
$$
$$
+ \left[e^{-i E_1 t} - e^{-i E_2 t} \right] sin\theta cos \theta
\nu_\mu (0) .
$$
\par
The probability that neutrino $\nu_e$ produced at moment $t = 0$
will be at moment $t \neq 0$ in the state of $\nu_\mu$ neutrino,
is given by a squared absolute value of the amplitude in (38)
$$
P(\nu_e \rightarrow \nu_\mu) = \mid(\nu_\mu(0) \cdot
\nu_e(t))\mid^2 = \eqno(39)
$$
$$
 = {1\over 2} \sin^{2} 2\theta \left[1 - cos ((E_2 - E_1) t)
\right] = \sin^{2}\, (2 \theta) \sin^2\, ((m^{2}_{2} - m^{2}_{1})
/ 2p) t.
$$
\par
The probability of $\nu_e \to \nu_e$ is given by the following
expression:
$$
P(\nu_e \rightarrow \nu_e) = 1 -  \sin^{2}\, (2 \theta) \sin^2\,
((m^{2}_{2} - m^{2}_{1}) / 2p) t, \eqno(40)
$$
where $p^2 \gg m^2_1, m^2_2$, and $E_1 = \sqrt{p^2 + m_1^2} \simeq
p + \frac{m^2_1}{2 p}$, $E_2 \simeq p + \frac{m^2_2}{2 p}$,
\par
Then the length $L_{1 2}$ of $\nu_e, \nu_\mu$ neutrino
oscillations is
$$
L_{1 2} = 2\pi  {2p \over {\mid m^{2}_{2} - m^{2}_{1} \mid}},
\eqno(41)
$$
or
$$
L_{1 2} = 2\pi  {2p \over {(m_{\nu_\mu} + m_{\nu_e})
\sqrt{(m_{\nu_\mu} - m_{\nu_e})^2 + (2 m_{\nu_e \nu_\mu})^2} }}.
\eqno(41')
$$
Now let us consider the cases of $\nu_\mu \to \nu_\tau$ and $\nu_e
\to \nu_\tau$ mixings and oscillations. We will give final
expressions without a detailed consideration.

\subsection{The case of $\nu_e, \nu_\tau$ neutrino oscillations}

\par
Nondiagonal mass matrix of $\nu_e, \nu_\tau$ neutrinos has the
following form:
$$
\left(\begin{array}{cc} m_{\nu_e} & m_{\nu_e \nu_\tau} \\
m_{\nu_e \nu_\tau}  & m_{\nu_\tau}
\end{array} \right) .  \eqno(42)
$$
After diagonalizing this matrix by turning on angle $\beta$
($\beta \equiv \theta_{1 3}$) we get
$$
U^{-1} \left(\begin{array}{cc} m_{\nu_e} & m_{\nu_e \nu_\tau} \\
m_{\nu_e \nu_\tau}  & m_{\nu_\tau} \end{array} \right) U
= \left(\begin{array}{cc} m_{\nu_1} & 0 \\
0  & m_{\nu_3} \end{array} \right), \quad U = \left(\begin{array}{cc} cos\beta & -sin\beta \\
sin\beta & cos\beta \end{array} \right) .
$$
$$
tg 2\beta = \frac{2m_{\nu_e \nu_\tau}}{\mid m_{\nu_e} -
m_{\nu_\tau} \mid}   ,
$$
$$
sin 2\beta = \frac{2m_{\nu_e \nu_\tau}}{\sqrt{(m_{\nu_e} -
m_{\nu_\tau})^2 +(2m_{\nu_e \nu_\tau})^2}}  ,  \eqno(43)
$$
$$
m_{1, 3} \equiv  m_{\nu_1 \nu_3} = {1\over 2} \left[ (m_{\nu_e} +
m_{\nu_\tau}) \pm \left((m_{\nu_e} - m_{\nu_\tau})^2 + 4
m^{2}_{\nu_e \nu_\tau} \right)^{1/2} \right]  , \eqno(44)
$$
where $\nu_1$ and $\nu_3$ states are eigenstates of the weak
interactions violating lepton numbers. Now these states are
superposition states of $\nu_e, \nu_\tau$ neutrinos:
$$
\begin{array} {cc} \nu_1 = {cos \beta \nu_e - sin \beta \nu_\tau} , \\ \nu_3
= { sin \beta \nu_e + cos \beta \nu_\tau}, \end{array} \eqno(45)
$$
and the inverse transformation gives:
$$
\begin{array} {cc} \nu_e = {cos \beta \nu_1 + sin \beta \nu_3} , \\ \nu_\tau
= {- sin \beta \nu_1 + cos \beta \nu_3} ,\end{array} . \eqno(46)
$$
Now the mass Lagrangian of two neutrinos ($\nu_e, \nu_\tau$) has
the following form:
$$
\begin{array}{c}{\cal L}_{M} = -  \frac{1}{2} \left[m_{\nu_e}
\bar \nu_e \nu_e + m_{\nu_\tau} \bar \nu_{\tau} \nu_{\tau } +
m_{\nu_e \nu_{\tau}}(\bar \nu_e \nu_{\tau } + \bar \nu_{\tau }
\nu_e) \right] \equiv \\
\equiv  -  \frac{1}{2} (\bar \nu_e, \bar \nu_\tau)
\left(\begin{array}{cc} m_{\nu_e} & m_{\nu_e \nu_{\tau }} \\
m_{\nu_{\mu} \nu_\tau} & m_{\nu_\tau} \end{array} \right)
\left(\begin{array}{c} \nu_e \\ \nu_{\tau } \end{array} \right)\to
\end{array}
\eqno(47)
$$
$$
\to -  \frac{1}{2} \left[ m_1 \bar \nu_1 \nu_1 + m_3 \bar \nu_3
\nu_3 \right],
$$
\par
Expression for $sin^2 2\beta$ has the following form:
$$
\sin^2\, (2 \beta) = \frac{(2m_{\nu_{e} \nu_{\mu}})^2} {(m_{\nu_e}
- m_{\nu_\tau})^2 +(2m_{\nu_e \nu_{\mu}})^2}. \eqno(48)
$$
\par
The probability that neutrino $\nu_e$ produced at moment $t = 0$
will be at moment $t \neq 0$ in the state of $\nu_\tau$ neutrino,
is given by the following expression:
$$
\begin{array}{c}
P(\nu_e \rightarrow \nu_\tau) = \mid(\nu_\tau(0) \cdot \nu_e(t))\mid^2 =\\
 = {1\over 2} \sin^{2} (2\beta) \left[1 - cos ((E_3 - E_1) t)
\right] \end{array} \eqno(49)
$$
$$
= \sin^{2}\, (2 \beta) \sin^2\, ((m^2_3 - m^2_1) / 2p) t.
$$
\par
The probability of $\nu_e \to \nu_e$ is given by the following
expression:
$$
P(\nu_e \rightarrow \nu_e) = 1 -  \sin^{2}\, (2 \beta) \sin^2\,
((m^2_3 - m^2_1) / 2p) t, \eqno(50)
$$
where $p^2 \gg m^2_1, m^2_3$, and $E_1 = \sqrt{p^2 + m_1^2} \simeq
p + \frac{m^2_1}{2 p}$, $E_3 \simeq p + \frac{m^2_3}{2 p}$,
\par
Then the length $L_{1 3}$ of $\nu_e, \nu_\tau$ neutrino
oscillations is
$$
L_{1 3} = 2\pi  {2p \over {\mid m^2_3 - m^2_1 \mid}}, \eqno(51)
$$
or
$$
L_{1 3} = 2\pi  {2p \over{(m_{\nu_\tau} + m_{\nu_e})
\sqrt{(m_{\nu_\tau} - m_{\nu_e})^2 + (2 m_{\nu_e \nu_\tau})^2} }}.
\eqno(51')
$$

\subsection{The case of $\nu_\mu, \nu_\tau$ neutrino oscillations}

\par
The nondiagonal mass matrix of $\nu_\mu, \nu_\tau$ neutrinos has
the following form:
$$
\left(\begin{array}{cc} m_{\nu_\mu} & m_{\nu_\mu \nu_\tau} \\
m_{\nu_\mu \nu_\tau}  & m_{\nu_\tau}
\end{array} \right) .  \eqno(52)
$$
After diagonalizing this matrix by turning on angle $\gamma$
($\gamma \equiv \theta_{2 3}$) we get
$$
U^{-1} \left(\begin{array}{cc} m_{\nu_\mu} & m_{\nu_\mu \nu_\tau} \\
m_{\nu_\mu \nu_\tau}  & m_{\nu_\tau} \end{array} \right) U
= \left(\begin{array}{cc} m_{\nu_2} & 0 \\
0  & m_{\nu_3} \end{array} \right), \quad U = \left(\begin{array}{cc} cos\gamma & -sin\gamma \\
sin\gamma & cos\gamma \end{array} \right) .
$$
$$
tg 2\gamma = \frac{2m_{\nu_\mu \nu_\tau}}{\mid m_{\nu_\mu} -
m_{\nu_\tau} \mid}   ,
$$
$$
sin 2\gamma = \frac{2m_{\nu_\mu \nu_\tau}}{\sqrt{(m_{\nu_\mu} -
m_{\nu_\tau})^2 +(2m_{\nu_\mu \nu_\tau})^2}}  ,  \eqno(53)
$$
$$
m_{2, 3} \equiv  m_{\nu_2 \nu_3} = {1\over 2} \left[ (m_{\nu_\mu}
+ m_{\nu_\tau}) \pm \left((m_{\nu_\mu} - m_{\nu_\tau})^2 + 4
m^{2}_{\nu_\mu \nu_\tau} \right)^{1/2} \right]  , \eqno(54)
$$
where $\nu_2$ and $\nu_3$ states are eigenstates of the weak
interactions violating lepton numbers. Now these states are
superposition states of $\nu_\mu, \nu_\tau$ neutrinos:
$$
\begin{array} {cc} \nu_2 = {cos \gamma \nu_\mu - sin \gamma \nu_\tau} , \\ \nu_3
= { sin \gamma \nu_\mu + cos \gamma \nu_\tau}, \end{array}
\eqno(55)
$$
and the inverse transformation gives:
$$
\begin{array} {cc} \nu_\mu = {cos \gamma \nu_2 + sin \gamma \nu_3} , \\ \nu_\tau
= {- sin \gamma \nu_2 + cos \gamma \nu_3} . \end{array}  \eqno(56)
$$
Now the mass Lagrangian of two neutrinos ($\nu_\mu, \nu_\tau$) has
the following form:
$$
\begin{array}{c}{\cal L}_{M} = -  \frac{1}{2} \left[m_{\nu_\mu}
\bar \nu_\mu \nu_\mu + m_{\nu_\tau} \bar \nu_{\tau} \nu_{\tau } +
m_{\nu_\mu \nu_{\tau}}(\bar \nu_\mu \nu_{\tau } + \bar \nu_{\tau }
\nu_\mu) \right] \equiv \\
\equiv  -  \frac{1}{2} (\bar \nu_\mu, \bar \nu_\tau)
\left(\begin{array}{cc} m_{\nu_\mu} & m_{\nu_\mu \nu_{\tau }} \\
m_{\nu_{\mu} \nu_\tau} & m_{\nu_\tau} \end{array} \right)
\left(\begin{array}{c} \nu_\mu \\ \nu_{\tau } \end{array}
\right)\to
\end{array}
\eqno(57)
$$
$$
\to -  \frac{1}{2} \left[ m_2 \bar \nu_2 \nu_2 + m_3 \bar \nu_3
\nu_3 \right],
$$
\par
Expression for $sin^2 2\gamma$ has the following form:
$$
\sin^2\, (2 \gamma) = \frac{(2m_{\nu_{e} \nu_{\mu}})^2}
{(m_{\nu_\mu} - m_{\nu_\tau})^2 +(2m_{\nu_\mu \nu_{\mu}})^2}.
\eqno(58)
$$
\par
The probability that neutrino $\nu_\mu$ produced at moment $t = 0$
will be at moment $t \neq 0$ in the state of $\nu_\tau$ neutrino,
is given by the following expression:
$$
\begin{array}{c}
P(\nu_\mu \rightarrow \nu_\tau) = \mid(\nu_\tau(0) \cdot \nu_\mu(t))\mid^2 =\\
 = {1\over 2} \sin^{2} (2\gamma) \left[1 - cos ((E_3 - E_2) t)
\right] \end{array} \eqno(59)
$$
$$
= \sin^{2}\, (2 \gamma) \sin^2\, ((m^2_3 - m^2_2) / 2p) t.
$$
\par
The probability of $\nu_\mu \to \nu_\mu$ is given by the following
expression:
$$
P(\nu_\mu \rightarrow \nu_\mu) = 1 -  \sin^{2}\, (2 \gamma)
\sin^2\, ((m^2_3 - m^2_2) / 2p) t, \eqno(60)
$$
where $p^2 \gg m^2_2, m^2_3$, and $E_2 = \sqrt{p^2 + m_2^2} \simeq
p + \frac{m^2_2}{2 p}$, $E_3 \simeq p + \frac{m^2_3}{2 p}$,
\par
Then the $L_{1 2}$ of $\nu_\mu, \nu_\tau$ neutrino oscillations
length is:
$$
L_{2 3} = 2\pi  {2p \over {\mid m^2_3 - m^2_2 \mid}}, \eqno(61)
$$
or
$$
L_{2 3} = 2\pi  {2p \over{(m_{\nu_\tau} + m_{\nu_\mu})
\sqrt{(m_{\nu_\tau} - m_{\nu_\mu})^2 + (2 m_{\nu_\mu \nu_\tau})^2}
}}. \eqno(61')
$$
Now we analyze the modern experimental data on neutrino
oscillations by using the expressions obtained in the scheme of
mass mixings.

\section{Analysis of the modern experimental data on neutrino
oscillations by using the results obtained in the scheme of mass
mixings}

\subsection{Analysis of $\nu_e, \nu_\mu$ processes}

The process of $\nu_e, \nu_\mu$ oscillations was studied in
experiment KamLAND \cite{11} and they obtained
$$
\theta \equiv \theta_{1 2} \simeq 34^{o}, \quad sin^2 (2 \theta_{1
2}) \simeq 0.860,
$$
$$
\mid m_2^2 - m_1^2 \mid = 7.50 (+ 0.19 -0.20) \times 10^{-5} eV^2.
\eqno(62)
$$
\par
Using expression (33) for $sin^2 (2 \theta)$,
$$
\sin^2 (2 \theta) \equiv \sin^2 (2 \theta_{1 2})=
\frac{(2m_{\nu_{e} \nu_{\mu}})^2} {(m_{\nu_e} - m_{\nu_\mu})^2
+(2m_{\nu_e \nu_{\mu}})^2} \simeq 0.860 , \eqno(63)
$$
we can do the following conclusion:
$$
(m_{\nu_e} - m_{\nu_\mu})^2 \simeq 0.163 (2m_{\nu_{e}
\nu_{\mu}})^2, \quad (2m_{\nu_{e} \nu_{\mu}})^2 \simeq 6.14
(m_{\nu_e} - m_{\nu_\mu})^2
$$
i.e., the difference between masses of $\nu_e$ and $\nu_\mu$
neutrinos is less than the nondiagonal mass term. Then deposit of
$\nu_e, \nu_\mu$ neutrino mass difference in the length of $\nu_e,
\nu_\mu$ neutrino oscillations is very small (see expr. (41')):
$$
L_{1 2} = 2\pi \frac{2p}{\mid m^2_2 - m^2_1 \mid} \equiv 2\pi
\frac{2p} {(m_{\nu_\mu} + m_{\nu_e}) \sqrt{(m_{\nu_\mu} -
m_{\nu_e})^2 + (2 m_{\nu_e \nu_\mu})^2}} \approx
$$
$$
\approx 2\pi \frac{2p}{(m_{\nu_\mu} + m_{\nu_e}) \sqrt{(2 m_{\nu_e
\nu_\mu})^2}} , \eqno(64)
$$
i.e., the length of $\nu_e, \nu_\mu$ neutrino oscillations is
mainly formed by the nondiagonal mass term $2 m_{\nu_e \nu_\mu}$.

\subsection{Analysis of $\nu_\mu, \nu_\tau$ processes}

The process of $\nu_e, \nu_\mu$ oscillations was studied in
experiment Super-Kamiokande \cite{15} and they obtained
$$
\gamma \equiv \theta_{2 3} \simeq 45^{o}, \quad sin^2 (2 \theta_{2
3}) \simeq 1.0 , \eqno(65)
$$
and $m_3^2 - m_2^2 = 2.1 \times 10^{-3} eV^2$.
\par
Using expression (62) for $sin^2 (2 \gamma)$:
$$
\sin^2 (2 \gamma) \equiv \sin^2 (2 \theta_{2 3})=
\frac{(2m_{\nu_\mu \nu_\tau})^2} {(m_{\nu_\mu} - m_{\nu_\tau})^2
+(2m_{\nu_\mu \nu_\tau})^2} \simeq 1.0 , \eqno(66)
$$
we can do the following conclusion:
$$
(m_{\nu_\mu} - m_{\nu_\tau})^2 \simeq 0.0, \quad m_{\nu_\mu}
\simeq m_{\nu_\tau},
$$
i.e., the mass of $\nu_\mu$ neutrino is about the $\nu_\tau$
neutrinos mass. Then the deposit of $\nu_\mu, \nu_\tau$ neutrino
mass difference in the length of $\nu_e, \nu_\mu$ neutrino
oscillations is about zero (see expression. (65), (61'))
$$
L_{2 3} = 2\pi \frac{2p}{\mid m^2_3 - m^2_2 \mid} \equiv 2\pi
\frac{2p} {(m_{\nu_\mu} + m_{\nu_\tau}) (2 m_{\nu_\mu \nu_\tau})}
= 2\pi \frac{2p} {(2 m_{\nu_\mu}) (2 m_{\nu_\mu \nu_\tau})} ,
\eqno(67)
$$
i.e., the length of $\nu_\mu, \nu_\tau$ neutrino oscillations is
mainly formed by the nondiagonal mass term $2 m_{\nu_\mu
\nu_\tau}$. It is interesting to remark that if to suppose that $2
m_{\nu_\mu} \approx 1 eV$ then from the expression $2 m_{\nu_\mu}
2 m_{\nu_\mu \nu_\tau} = 2.1 \times 10^{-3} eV^2$ we get
$m_{\nu_\mu \nu_\tau} \approx 10^{-3} eV$. This value is very big
in contrast to the $K^o$ meson oscillation case where the
analogous term $m_{K^o \bar K^o} \approx 10^{-6} eV$.

\subsection{Analysis of $\nu_e, \nu_\tau$ processes}

The process of $\nu_e, \nu_\mu$ oscillations was studied in
experiment KamLAND \cite{11} and they obtained
$$
\beta \equiv \theta_{1 3} \leq 13^{o}, \quad sin^2 (2 \theta_{1
3}) \leq 0.192, \eqno(68)
$$
$\mid m_3^2 - m_1^2 \mid$ is still unknown until now. Using
expression (68) for $sin^2 (2 \beta)$
$$
\sin^2 (2 \beta) \equiv \sin^2 (2 \theta_{1 3})=
\frac{(2m_{\nu_{e} \nu_{\tau}})^2} {(m_{\nu_e} - m_{\nu_\tau})^2
+(2m_{\nu_e \nu_{\tau}})^2} \simeq 0.192 , \eqno(69)
$$
we can do the following conclusion:
$$
(m_{\nu_e} - m_{\nu_\tau})^2 \simeq 4.21 (2m_{\nu_{e}
\nu_{\mu}})^2, \quad (2m_{\nu_{e} \nu_{\mu}})^2 \simeq 0.238
(m_{\nu_e} - m_{\nu_\tau})^2,
$$
i.e., the difference between the masses of $\nu_e$ and $\nu_\tau$
neutrinos is much more than the nondiagonal mass term. Then the
deposit of $\nu_e, \nu_\mu$ neutrino mass difference in the length
of $\nu_e, \nu_\tau$ neutrino oscillations are very big (see
expression. (51')):
$$
L_{1 3} = 2\pi \frac{2p}{\mid m^3_2 - m^2_1 \mid} \equiv 2\pi
\frac{2p} {(m_{\nu_\tau} + m_{\nu_e}) \sqrt{(m_{\nu_\tau} -
m_{\nu_e})^2 + (2 m_{\nu_e \nu_\tau})^2}} \approx
$$
$$
\approx 2\pi \frac{2p} {(m_{\nu_\tau} + m_{\nu_e})
\sqrt{(m_{\nu_\tau} - m_{\nu_e})^2 }} , \eqno(70)
$$
i.e., the length of $\nu_e, \nu_\tau$ neutrino oscillations is
mainly formed by the mass difference of these neutrinos.

\subsection{Remarks about the problem in neutrino oscillations
processes in the framework of the mass mixings scheme}

If angle mixings $\theta_{2 3}$ of $\nu_\mu, \nu_\tau$ at neutrino
oscillations are maximal, i.e., $\pi/4$ \cite{11}, then in the
framework of the mass mixings scheme the masses of $\nu_\mu,
\nu_\tau$ neutrinos have to be nearly equal, i.e., $m_\mu \simeq
m_\tau$. Farther in the framework of this approach the equality of
oscillation lengths of $L_{2 3}$ and $L_{1 3}$ is impossible to
obtain without additional supposition (see expressions (51'),
(61')), in contrast to the conventional approach, since the
nondiagonal mass components $m_{\nu_\mu \nu_\tau}$, $m_{\nu_e
\nu_\tau}$ of the mass matrix cannot be equal by definition. Then
the natural solution of this problem is to suppose that $L_{1 3}$
is larger than $L_{2 3}$, then the length of neutrino oscillations
$L_{1 3}$ has to be larger than $L_{2 3}$, i.e., the value of
$\theta_{1 3}$ is necessary to search on distances more than $L_{2
3}$.
\par
To solve this problem in the framework of the standard approach
\cite{10}, it is necessary to suppose that
$$
(m_2^2 - m_1^2) \neq (m_3^2 - m_1^2) - (m_3^2 - m_2^2). \eqno(71)
$$
Obviously, it is possible if to suppose that 4 neutrino
oscillations are realized instead of 3 neutrino oscillations,
i.e., if there is the fourth component.

\section{Conclusion}

On the example of $K^o$ mixings and oscillations we have
considered mixings and oscillations of $\nu_e, \nu_\mu, \nu_\tau$
neutrinos. The analysis of the obtained results has been done by
using modern experimental data on neutrino oscillations. In these
experimental data there is one problem. If we use the conventional
theoretical approach \cite{10}, then $ L_{1 3} \approx L_{2 3}$
($L_{1 2} \gg L_{2 3}$) and the mixing angle of $\theta_{1 3}$ is
very small. However, angle mixings $\theta_{2 3}, \theta_{1 2}$
are big and close to maximal angle $\pi/4$. The problem is: why is
mixing angle $\theta_{1 3}$ so small?

Since the angle mixings $\theta_{2 3}$ of $\nu_\mu, \nu_\tau$ at
neutrino oscillations is maximal \cite{12}, i.e., $\pi/4$ then in
the framework of the mass mixings scheme the masses of $\nu_\mu,
\nu_\tau$ neutrinos have to be nearly equal, i.e., $m_\mu \simeq
m_\tau$. In the framework of this approach the equality of
oscillation lengths of $L_{2 3}$ and $L_{1 3}$ is impossible to
obtain without additional supposition (see expressions (51'),
(61')) since the nondiagonal mass components $m_{\nu_\mu
\nu_\tau}$, $m_{\nu_e \nu_\tau}$ of the mass matrix cannot be
equal by definition. Then the natural solution of this problem is
to suppose that $L_{1 3}$ is larger than $L_{2 3}$ ($L_{1 3}
> L_{2 3}$). Then it is necessary to examine $\nu_e, \nu_\tau$
neutrino oscillations at much longer distances than $L_{2 3}$ to
search for the value of $\theta_{1 3}$.
\par
To solve this problem in the framework of the standard approach
\cite{10}, it is necessary to suppose that $(m_2^2 - m_1^2) \neq
(m_3^2 - m_1^2) - (m_3^2 - m_2^2)$. Obviously, it is possible if
to suppose that 4 neutrino oscillations are realized instead of 3
neutrino oscillations, i.e., if there is the fourth component.


\end{document}